\documentstyle{mn}

\newcommand{\HI}{\hbox{H\,{\sc i} }}
\newcommand\sfrac[2]{{\textstyle{\frac{#1}{#2}}}}
\newcommand{\cm}{\,{\rm cm}}

\newcommand{\cmcube}{\,{\rm cm^{-3}}}

\newcommand{\cms}{\,{\rm cm\,s^{-2}}}
\newcommand{\kms}{\,{\rm km\,s^{-1}}}

\newcommand{\kpc}{\,{\rm kpc}}

\newcommand{\mkG}{\,\mu{\rm G}}

\newcommand{\p}{\,{\rm pc}}
\newcommand{\radm}{\,{\rm rad\,m^{-2}}}

\newcommand{\yr}{\,{\rm yr}}

\input epsf
\epsfverbosetrue

\title[Hydrostatic equilibrium in a magnetized, warped Galactic
disc]{Hydrostatic equilibrium in a magnetized, warped Galactic disc}

\author[A.\ Fletcher and A.\ Shukurov]{Andrew Fletcher and
Anvar Shukurov\thanks{E-mails: {\tt andrew.fletcher@ncl.ac.uk} and
        {\tt anvar.shukurov@ncl.ac.uk}}\\
  Department of Mathematics, University of Newcastle upon Tyne, NE1
  7RU}

\date{Accepted .. ..... 2001.
      Received .. ..... 2001;
      in original form .. ..... 2000}

\pagerange{\pageref{firstpage}--\pageref{lastpage}}

\pubyear{2001}

\begin{document}

\maketitle

\label{firstpage}

%********************
% Abstract & Keywords
%********************

\begin{abstract}
  Hydrostatic equilibrium of the multiphase interstellar medium in the
  solar vicinity is reconsidered, with the regular and turbulent
  magnetic fields treated separately. The regular magnetic field
  strength required to support the gas is consistent with independent
  estimates provided energy equipartition is maintained between
  turbulence and random magnetic fields. Our results indicate that a
  midplane value of $B_{0}=4\mkG$ for the regular magnetic field near
  the Sun leads to more attractive models than $B_{0}=2\mkG$. The
  vertical profiles of both the regular and random magnetic fields
  contain disc and halo components whose parameters we have
  determined. The layer at $1\la|z|\la4\kpc$ can be overpressured and
  an outflow at a speed of about $50\kms$ may occur there, presumably
  associated with a Galactic fountain flow, if $B_{0}\simeq 2\mkG$.

  We show that hydrostatic equilibrium in a warped disc must produce
  asymmetric density distributions in $z$, in rough agreement with \HI
  observations in the outer Galaxy. This asymmetry may be a useful
  diagnostic of the details of the warping mechanism in the Milky Way
  and other galaxies. We find indications that gas and magnetic field
  pressures are different above and below the warped midplane in the
  outer Galaxy and quantify the difference in terms of turbulent
  velocity and/or magnetic field strength.
\end{abstract}


% ********************
\section{Introduction}
% ********************
Models assuming hydrostatic equilibrium in the Galactic gas layer are
often used to describe the global state of the interstellar medium
(ISM) (Parker 1966, Badhwar \& Stephens 1977, Bloemen 1987, Boulares
\& Cox 1990, Lockman \& Gehman 1991, Kalberla \& Kerp 1998). Despite
matter outflows in the form of Galactic fountain (Shapiro \& Field
1976, Bregman 1980, Habe \& Ikeuchi 1980) and chimneys (Norman \&
Ikeuchi 1989), the assumption of hydrostatic equilibrium appears to be
viable when considering the average state of the ISM over large scales
or times.  Korpi et al.\ (1999) argue that the warm ISM is on average
in a state of equilibrium, although not the hot phase within
$\simeq1\kpc$ from the midplane.

Magnetic fields play an important role in the pressure balance of the
ISM (e.g.\ Parker 1979, Boulares \& Cox 1990). Both theory and
observations of interstellar magnetic fields assert that magnetic
energy density should be comparable to turbulent and thermal energy
densities of the gas (e.g.\ Zweibel \& McKee 1995, Beck et al.\ 1996),
with a direct implication for a similar balance in the pressures.

The Galactic magnetic field has two distinct components, the regular
field $B$ on length scales of the order 1\,kpc and the turbulent field
$b$ with a typical scale of order 100\,pc. Galactic dynamo theory
predicts that the strength of the regular field depends on magnetic
diffusivity, rotational shear, the kinetic energy density of the ISM,
etc.\ (Ruzmaikin, Sokoloff \& Shukurov 1988, Ch.~VII; Shukurov 1998).
The turbulent magnetic field strength probably scales simply with the
kinetic energy density of the turbulent gas (Beck et al.\ 1996,
Subramanian 1999). Using the simplest form of this scaling in a model
of the hydrostatic support of the ISM in the solar vicinity allows the
pressure support from $B$ and $b$ to be separately quantified. In this
paper we discuss implications, for the strength and vertical
distribution, of both regular and turbulent magnetic fields, resulting
from hydrostatic equilibrium models. The validity of the model is
assessed using independent information on the magnetic fields in the
ISM.

Descriptions of the state of the interstellar medium tend to
concentrate on the situation in the solar vicinity. However, the gas
disc is warped in the outer Galaxy (Binney 1992), and the warping
implies that there is an additional force involved in the vertical
equilibrium balance. In Section~\ref{OG} we use simple arguments to
illustrate some effects of the warp on the vertical gas distribution
and to suggest that the vertical gas distribution can have a
characteristic asymmetry in the warped disc. Because of the warp, gas
densities at $\pm0.5\kpc$ above and below the displaced midplane can
differ by a factor of $1.2$. An asymmetry of this kind has been
observed in the outer Galaxy (Diplas \& Savage 1991), but went
unnoticed.

%***********************************************************
\section{Hydrostatic equilibrium in the solar neighbourhood} \label{HE}
%***********************************************************
%**********
\subsection{A model of hydrostatic support}
%**********
We assume that, on average, the gas at the solar position is in a
state of vertical hydrostatic equilibrium,
\begin{equation}
  \label{eq:equilibrium}
  \frac{\partial P}{\partial z}=\rho g\;,
\end{equation}
where $P$ and $\rho$ are the total pressure and gas density
respectively, $g$ is the gravitational acceleration and $z$ is the
height above the midplane.  The assumption is probably weakest when
considering the hot, ionized gas which presumably originates in the
disc but fills the Galactic halo, and so is involved in systematic
vertical motions.  However, hydrostatic equilibrium is still a
plausible expectation for large areas, of order a few kiloparsecs in
size.

The equilibrium model includes several contributions to the total
pressure,
\[
P=P_{\rm g}+P_{\rm m}+P_{\rm cr}\;,
\]
with
\[
P_{\rm g}(z)=\sum_{i}\rho_{i}v_{zi}^2 + \sum_{i}n_{i}kT_{i}
\]
the gas pressure consisting of the turbulent and thermal components
(with contributions from individual phases of the ISM labelled with
index $i$), $P_{\rm m}$ the magnetic pressure due to the regular, $B$,
and random, $b$, magnetic field components, and P$_{\rm cr}$ the
cosmic ray pressure. Here $v_{zi}$ is the vertical component of
the rms turbulent velocity in the {\it{i}}'th phase, $n_i$ the gas
number density, $T_i$ the temperature and $k$ is Boltzmann's constant.

We separate the pressure due to the regular and turbulent magnetic
fields by assuming that the energy density of the turbulent magnetic
field is a multiple of the kinetic energy density of the turbulent
gas,
\begin{equation}
  \label{eq:energy}
P_{\rm m}= \frac{B^2+b^2}{8\pi}\;,\qquad \frac{b^2}{8\pi}=\sfrac{1}{2}K\sum_{i}\rho_{i}
v_{i}^2\;,
\end{equation}
with $K$ a constant and $v_{i}^2 = 3v_{zi}^2$ is the total turbulent
velocity. A value of $K\simeq 1$ is in line with theoretical arguments
showing that equipartition should exist between turbulent kinetic
energy and magnetic energy densities (Zweibel \& McKee 1995, Beck et
al.\ 1996).  If the turbulent magnetic field is concentrated into flux
ropes and within these ropes energy equipartition also exists
(Subramanian 1999), then $K$ is the filling factor of the ropes.

By taking $\bmath{B}$ to be parallel to the Galactic plane we neglect
the contribution of magnetic tension due to the regular field to the
support of the ISM (cf.\ Boulares \& Cox 1990). The vertical regular
magnetic field is definitely negligible near the midplane away from
the disc centre (Ruzmaikin et al.\  1988, Beck et al.\ 1996). The
dominance of the horizontal magnetic field is more questionable in the
Galactic halo. Significant vertical magnetic fields have been detected
in the haloes of only two galaxies with strong star formation (NGC
4631, Hummel et al.\ 1988, 1991 and M82, Reuter et al.\ 1994), but the
Milky Way hardly belongs to this type.  Vertical dust filaments
abundant in many galaxies in the disc-halo interface (Sofue 1987) may
indicate significant localized vertical magnetic fields there.  These
can be a result of shearing the horizontal field in the disc by a
galactic fountain and/or chimneys (cf. Korpi et al.\ 1999; see also
Elstner et al.\ 1995).  Still, this interpretation is compatible with
a predominantly horizontal regular magnetic field in the main parts
of both the disc and the halo away from the interface (e.g.\ in M51
-- Berkhuijsen et al.\ 1997).  In the absence of direct observational
evidence for the halo of the Milky Way, we refer to numerous
theoretical models which invariably indicate the dominance of the
horizontal regular magnetic field under a variety of physical setups
(e.g.\ Brandenburg et al.\ 1992, 1993; Brandenburg, Moss \& Shukurov
1995).

% *** TABLE ----------------------------------------------------------
\begin{table}
\caption{Gas components of the modelled ISM at the solar position}
\begin{tabular}{@{}lcccc}
  %\hline
  Component &$n_{0}$ &$h$ &$T$ &$v_{z}$ \\
  &  cm$^{-3}$ &pc &K &km\,s$^{-1}$ \\[3pt]
  %\hline
Molecular$^{\rm (a)}$    & 0.3              &70  &20            &4.5\\
Cold$^{\rm (a)}$         & 0.3              &135 &100            &6\\
Warm$^{\rm (a)}$         & 0.1              &400 &$8\times10^{3}$&9\\
Diffuse ionized$^{\rm(b)}$&$2.4\times10^{-2}$&950&$8\times10^{3}$&12\\
Halo, ionized$^{\rm(c)}$ & $1.2\times10^{-3}$ & 4400
                                &$1.5\times10^{6}$ & $60$ \\
Halo, neutral$^{\rm(d)}$ & $1.3\times10^{-3}$ & 4400 & $10^{4}$ & $60$\\[3pt]
%\hline
\end{tabular}
Midplane number density $n_{0}$, scale height $h$, temperature $T$ and
one-dimensional turbulent velocity dispersion $v_{z}$ for the ISM components
are given at the solar position.
All components are assumed to have a Gaussian distribution
except the halo gas, which is taken to have an exponential distribution.
References:
(a)~Bloemen \shortcite{blo},
(b)~Reynolds \shortcite{rey},
(c)~Pietz et al.\ \shortcite{pietal},
(d)~Kalberla et al.\ \shortcite{kaletal}.
\label{table1}
\end{table}
% *** TABLE ----------------------------------------------------------

The ISM has a multicomponent structure; parameters of the basic phases
are given in Table~\ref{table1}. Molecular hydrogen, together with the
cold and warm neutral phases make up the bulk of the gaseous disc
mass, the other components being the diffuse ionized gas (also known
as the Reynolds layer), a neutral halo and ionized halo gas
(Ferri\`ere 1998, Kalberla \& Kerp 1998). Different phases do not
necessarily have a distinct physical nature and the distinction is
often merely conventional; for example, neutral and ionized halo
components plausibly represent a single physical entity; likewise, the
separation of the warm and diffuse ionized phases may simply describe
the increase of ionization degree with height. The halo components are
present in the disc as the hot phase. The filling factor for them was
taken to be that of the hot phase, $0.3$ in the midplane, rising to
0.9 at $z=4\kpc$ and 1 further up in the halo, in an approximation to
Fig.~10 of Ferri\`ere (1995) (see also Rosen \& Bregman 1995, Korpi et
al.\ 1999). To allow for helium, nine percent by number of hydrogen
atoms has been added to obtain the mass density in each phase, and we
assume that the helium is fully ionized in the ionized halo component.

We note that observational estimates of the total random velocity in
the diffuse ionized gas yield supersonic values $20\kms$ (Reynolds
1990a).  This can be due to systematic (albeit random) upward motions
of the hot gas forming the base of the galactic fountain (cf.\ Korpi
et al.\ 1999) rather than to genuine turbulent motions; the ram
pressure of these motions contributes to the vertical support of the
gas and we can include it in the turbulent pressure.

The pressure of cosmic rays is assumed to be equal to that of the
total magnetic field, $P_{\rm cr}=P_{\rm m}$ where $P_{\rm m}$ is
given by Eq.~(\ref{eq:energy}). Altogether, the total pressure is
given by
\begin{equation} \label{eq:pressure}
P=(3K+1)\sum_i\rho_iv_{zi}^2+\sum_i n_i kT_i + \frac{B^2}{4\pi}\;.
\end{equation}

%**********
\subsection{Observational constraints on magnetic field near the Sun}
%**********

The vertical distribution of $B$ is poorly known, and we derive it in
Section~\ref{RMF} from Eqs~(\ref{eq:equilibrium}) and
(\ref{eq:pressure}), and then assess the validity of our results using
the following reasonably well established constraints.

First, if equipartition between the energy densities of cosmic rays
and magnetic fields is assumed then the total interstellar magnetic
field, $B_{\rm tot}=\sqrt{B^2+b^2}$, at $z=0$ is estimated from
synchrotron intensity as $B_{\rm tot}\simeq6\mkG$ (Beck 2001). With
observations of synchrotron polarization showing $B_0/B_{\rm
tot}\simeq0.6$, as described below, this gives $B_0\simeq4\mkG$ (Beck
2001); if anisotropic turbulent magnetic fields are present this is an
upper limit on $B_0$. Observations of Faraday rotation measures from
pulsars and extragalactic radio sources yield a midplane value of $B$
of $B_0\simeq2\mkG$ (Rand \& Kulkarni 1989, Lyne \& Smith 1989, Rand
\& Lyne 1994, Frick et al.\ 2001). Note that $B_0$ can be
significantly larger at a distance of a few kiloparsecs from the Sun
because the Sun is located close to a reversal of the large-scale
magnetic field (e.g.\ Beck et al.\ 1996). Magnetic field strengths
obtained from rotation measures can be underestimated if the ISM is in
pressure balance, so $B$ and $\rho$ are anti-correlated (Beck 2001).
Therefore we consider models with $B_0$ equal to $4\mkG$ and $2\mkG$.

Second, the ratio of the turbulent to regular magnetic field, $b/B$,
is directly related to the degree of polarization, ${\cal P}$, of
the Galactic synchrotron emission via ${\cal P}={\cal
P}_0B^2/(b^2+B^2)$ with ${\cal P}_0\approx0.7$, provided any
depolarization effects can be neglected (Burn 1966, Sokoloff et al.\
1998). Berkhuijsen (1971) found $b/B\simeq1.2$ and Philipps et al.\
(1981) argue that $b/B$ cannot be much larger than unity. Observations
of the Galactic polarized synchrotron background yield $1\la b/B\la3$
(Spoelstra 1984).  This is corroborated by analyses of Faraday
rotation measures of extragalactic radio sources and pulsars which
imply $b/B\simeq1.7$ (Ruzmaikin et al.\ 1988, Ch.\ IV and references
therein) and $b/B\simeq3$ (Rand \& Kulkarni 1989) (these are upper
limits if $B$ is underestimated).

Finally the vertical distribution of the total interstellar magnetic
field, can be obtained from the distribution of Galactic synchrotron
emission (Phillips et al.\ 1981, Beuermann, Kanbach \& Berkhuijsen
1985). The model used by Phillips et al.\ \shortcite{phetal} requires
halo emissivity of about 10 percent of that in the disc, extending to
about 10\,kpc beyond the midplane.  Roughly, emissivity is
proportional to the total magnetic field squared or to a higher power
if detailed equipartition between magnetic fields and cosmic rays is
maintained (Beck et al.\ 1996, Zweibel \& Heiles 1997, Sokoloff et
al.\ 1998). So the total magnetic field, $B_{\rm tot}$, in the halo is
expected to be approximately $1/3$--$1/2$ of that in the disc, $B_{\rm
  tot}(\rm{halo})/B_{\rm tot}(\rm{disc})=1/3$--$1/2$.

%********************************************************************
\section{The role of magnetic field in the support of the gas layer}
\label{RMF}
%********************************************************************
The vertical distribution of the interstellar magnetic field is even
more difficult to obtain from observations than that of the gas
components of the ISM. Therefore we obtain $B$ as a function of $z$
from a hydrostatic equilibrium model and then compare the result with
independent information briefly summarized above. This also allows us
to reassess the validity of the equilibrium model itself.

Equation~(\ref{eq:equilibrium}), with $P$ from
Eq.~(\ref{eq:pressure}), is solved for $B(z)$ using different values
of $K$. We use $g$ for the solar vicinity derived by Ferri\`ere (1998)
from Kuijken \& Gilmore (1989a,b),
\begin{equation}
  -g=A_1\frac{z}{(z^2+Z_1^2)^{1/2}}+A_2\frac{z}{Z_2}\;,
  \label{gz}
\end{equation}
where $A_1=4.4\times10^{-9}\cms,\ A_2=1.7\times10^{-9}\cms,\
Z_1=0.2\kpc$ and $Z_2=1\kpc$.

\subsection{Model with $B_0\simeq4\mkG$} \label{B4}
%--------------------------------------------------------------------
%***FIGURE---------------------------------------------------------
\begin{figure}
\epsfxsize=8.0cm
\centerline{\epsfbox{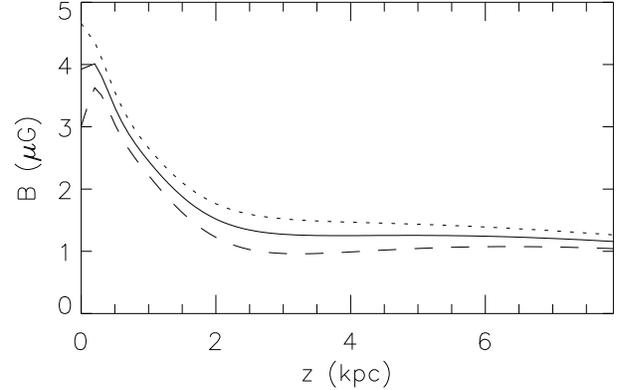}}
  \caption[]{The regular magnetic field, $B$, required for precise
  hydrostatic equilibrium with a midplane strength $B_0\simeq4\mkG$,
  for $K=0.8$ (solid), $K=0.6$ (dotted) and $K=1.0$ (dashed), where
  $K$ is the ratio of turbulent magnetic to kinetic pressures defined
  in Eq.~(\ref{eq:energy}).}  \label{fig:NewB}
\end{figure}
%***----------------------------------------------------------------

%***FIGURE---------------------------------------------------------
\begin{figure}
\epsfxsize=8.0cm
\centerline{\epsfbox{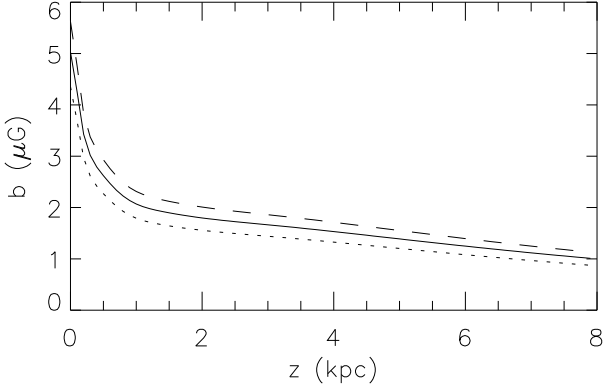}}
  \caption[]{Turbulent magnetic field strength, $b$, for the model of
    Fig.~\protect\ref{fig:NewB} (with $B_0\simeq4\mkG$) for $K=0.8$
    (solid), $K=0.6$ (dotted) and $K=1.0$ (dashed).}
    \label{fig:Newb}
\end{figure}
%***----------------------------------------------------------------

Constraining the midplane regular magnetic field to $B_0\simeq4\mkG$,
as suggested by cosmic ray energy equipartition arguments (Beck 2001),
we obtain the vertical distributions of $B$ and $b$ shown in
Figs~\ref{fig:NewB} and \ref{fig:Newb}. The best fit to all three
constraints described in section Section~\ref{HE} is with $K=0.8$, in
line with theoretical equipartition arguments (Zweibel \& McKee 1995,
Beck et al.\ 1996); we obtain $B_0 = 4\mkG$, $b/B = 1.25$ at $z=0$ and
$B_{\rm tot}({\rm halo})/B_{\rm tot}({\rm disc})\simeq1/3$.

Increasing $K$ amplifies the contribution of the turbulent magnetic
field and cosmic rays to the equilibrium balance, according to Eq.
(\ref{eq:energy}). Near the midplane, the strong turbulent pressure
(kinetic and magnetic) created by molecular cloud motions is amplified
to such an extent that the required strength of the regular magnetic
field for hydrostatic equilibrium begins to drop as $K$ increases (see
the solid and dashed lines in Fig.~\ref{fig:NewB}) reaching zero for
$K=1.4$.

Figure \ref{fig:NewB} suggests a two-component model of $B(z)$, a disc
with a scale height of $h_B\simeq 1\kpc$ and a halo extending
$z\simeq10\kpc$ with a roughly constant field strength of $B\simeq
1\mkG$. The scale height of the diffuse ionized gas
(Table~\ref{table1}) is roughly the same as $h_B$. This is also
in good agreement with Han and Qiao (1994) who have estimated
$h_B\simeq 1.2 \kpc$ from Faraday rotation measures of extragalactic
radio sources. Polarized radio emission, indicating the presence of a
regular magnetic field, has been detected high in the haloes of the
edge-on spiral galaxies to $z\simeq4\kpc$ in NGC~891 and to
$z\simeq8\kpc$ in NGC~4631 (Hummel, Beck \& Dahlem 1991).

Figure \ref{fig:Newb} shows a similar two-component structure of the
turbulent magnetic field $b$; a disc with scale height of
$h_b\simeq0.5\kpc$ and a slowly varying field in an extended halo
(with $b$ decreasing by a factor of two at $z\approx7\kpc$), where
$B\simeq b$.  We note a significant difference between the scale
heights of the regular and turbulent magnetic fields,
$h_B/h_b\simeq2$ for the narrower components.  This may have
important implications for theories of galactic magnetic fields.

%-----------------------------------------------------------------
\subsection{Model with $B_0\simeq2\mkG$} \label{B2}

%***FIGURE---------------------------------------------------------
\begin{figure}
\epsfxsize=8.0cm
\centerline{\epsfbox{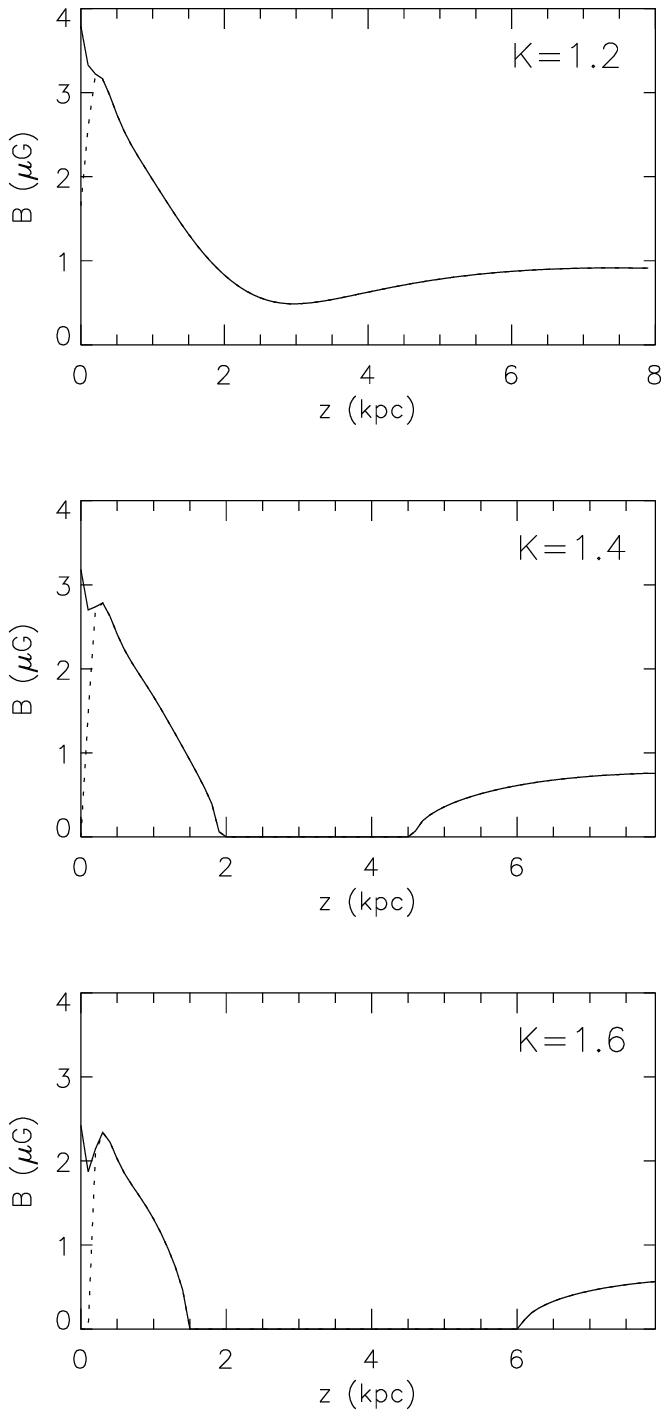}}
  \caption[]{The regular magnetic field, $B$, required for precise
  hydrostatic equilibrium with a midplane strength $B_0\simeq2\mkG$,
  for different values of $K$ with both the pressure and mass
  contributions from molecular clouds omitted (solid line) and with
  the molecular clouds included (dotted line up to $z\simeq 0.2 \kpc$,
  solid line above). In those ranges of $z$ where $B=0$, the
  hydrostatic equilibrium equations formally require negative magnetic
  pressure thereby indicating an overpressured region.} \label{fig:B}
\end{figure}
%***----------------------------------------------------------------
%
%***FIGURE---------------------------------------------------------

%-------------------------------------------
\begin{figure}
\epsfxsize=8.0cm
\centerline{\epsfbox{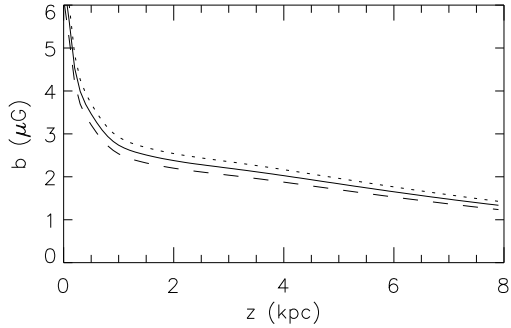}}
  \caption[]{Turbulent magnetic field strength, $b$ for the model with
    $B_0\simeq2\mkG$, excluding the effect of molecular cloud
    turbulent energy, for $K=1.2$ (dashed), $K=1.4$ (solid) and $K=1.6$
    (dotted).} \label{fig:b}
\end{figure}
%***----------------------------------------------------------------
%***FIGURE---------------------------------------------------------

%-------------------------------------------
\begin{figure}
\epsfxsize=8.0cm
\centerline{\epsfbox{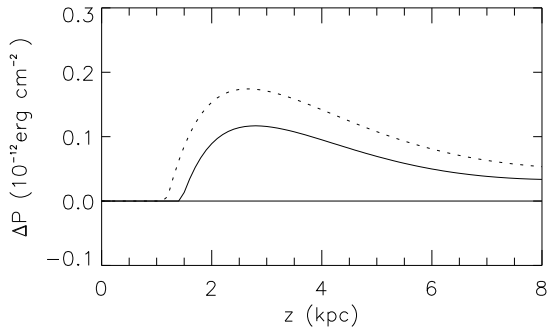}}
  \caption[]{Pressure excess over the hydrostatic equilibrium value,
  described by Eqs.~(\protect\ref{eq:pressure}) and
  (\protect\ref{eq:deltaP}), using $B$ shown in Fig.~\ref{fig:B} but
  with minimum of $B$ enforced to be $1\mkG$, for $K=1.4$ (solid) and
  $K=1.6$ (dashed).  Positive $\Delta P$ indicates overpressure.}
  \label{fig:P}
\end{figure}
%***----------------------------------------------------------------

If $B_0$ is significantly smaller than $4\mkG$ then the best fit to
the constraints described in Section~\ref{HE} is with $K=1.6$
(Figs~\ref{fig:B} and \ref{fig:b}); we obtain $B=2.4 \mkG$ at the
midplane (but excluding the molecular gas; see the discussion in
Section~\ref{mol}), $b/B\approx2.4$ and $B_{\rm total}({\rm
  halo})/B_{\rm total}({\rm disc})\approx1/3$. Realistic variations of
the gas parameters do not produce outcomes meeting all three criteria
far from $K=1.6$, which results in a constraint $1.5\la K\la1.7$.
Using a different representation of $g$ for the stellar disc derived
by Rohlfs (1977, p.35), with gravity from a spherical dark matter halo
added as $V_{\rm rot}^2 z/R_{\odot}$, where $V_{\rm rot}$ is the
rotation speed of the disc and $R_{\odot}$ the solar radius, the range
becomes $1.3\la K\la2.4$.  If $K$ is beyond this range, then either
$B_0>3 \mkG$, or $B_0<1 \mkG$, or $b_0/B_0>3$ must be accepted for the
midplane values.

%------------------------------------------------
\subsubsection{Overpressure above $z\simeq1$\,kpc} \label{op}
A strange feature of Fig.~\ref{fig:B}, for the better fits at
$K=1.4$ and $1.6$, is the apparent absence, or a strong decrease, of
$B$ at $z=1$--$6\kpc$. The reason for this is clear from
Fig.~\ref{fig:P}; where we show the total pressure excess over the
hydrostatic equilibrium value, 
\begin{equation}
  \Delta P=P-\int_{z}^{\infty}\rho g\,dz\;,
  \label{eq:deltaP}
\end{equation}
assuming $B(z)$ is as in Fig.~\ref{fig:B} but not less than $1\mkG$ at
any $z$. We see that there is more than enough pressure from the gas,
turbulent magnetic field and associated cosmic rays to support the
weight of the overlying ISM at $1\la z\la 6\kpc$. Any further
contribution from the regular magnetic field enhances overpressure in
this range of $z$.

The excess pressure at $z=1$--$4\kpc$, if real, apparently contributes
to driving the Galactic fountain flow. The vertical velocity driven
by the overpressure is $V_z\simeq(\sfrac12\Delta
P/\rho)^{1/2}\simeq50\kms$, significantly smaller than the
magnetosonic speed ($100\kms$ or more) and compatible with what can
be expected for galactic fountains. It is
notable that a similar systematic outflow of the hot gas occurs in
models of multiphase ISM driven by supernovae (Rosen \& Bregman 1995,
Korpi et al.\ 1999, Shukurov 1999).

%----------------------------------------------------------------
\subsubsection{Molecular clouds} \label{mol}
An unsatisfactory feature of the distribution shown with the dotted
line in Fig.~\ref{fig:B} (i.e.\ for a model including the molecular
gas) is the deep minimum of $B$ at $z=0$. This feature also occurs in
the model of Boulares \& Cox (1990, their Fig.~2). It is very unlikely
that $B$ can have such a sharp minimum even if it has a global odd
parity in $z$, resulting in the constraint $B_0=B(0)\equiv0$.  Apart
from increasing $B_0$ to $4\mkG$ (Section~3.1), a possible resolution
of this difficulty assumes that the coupling between kinetic energy
density and turbulent magnetic fields, parameterized by $K$, is weaker
for the bulk motion of molecular clouds than for other ISM phases.
This assumption is physically compelling because the clouds represent
the densest, self-gravitating part of the ISM, rather isolated from
the other diffuse phases. The minimum in $B$ does not occur if $K\la
0.5$ for the molecular clouds and $K\simeq 1$ in the other phases. If,
otherwise, $K\simeq 1$ for the bulk motion of molecular clouds as
well, then either $B$ must grow sharply in the range $|z|<100\p$ from
a small midplane value, or $B>3\mkG$ at $z=0$ (cf.\ the upper panel of
Fig.~\ref{fig:B}).

A possible interpretation is that the turbulent magnetic field
interacts differently with the molecular clouds than with the diffuse
phases of the ISM (e.g.\ because of the fluctuation dynamo action in
the diffuse gas -- see Sect.\ 4.1 in Beck et al.\ 1996 and
Subramanian 1999).  If the fractional volume swept by the molecular
clouds over the correlation time of $b$ is significantly less than
unity, only a correspondingly small fraction of the magnetic field
will be coupled to the clouds.  The number of clouds in a correlation
volume $l^{3}$ of the turbulent magnetic field is $N\simeq
(l/l_{c})^3 f_{c}$, where $f_{c}$ is the volume filling factor of the
clouds and $l_{\rm c}$ is the cloud size.  During the correlation
time of the magnetic field, $\tau\simeq l/v$, the clouds moving at a
speed $v_{\rm c}\approx8\kms$ (cf.  Table~\ref{table1}) sweep the
volume $V\simeq Nl_{\rm c}^2v_{\rm c}\tau$. So the clouds affect
magnetic fields within the fractional volume $V/l^3\simeq f_{\rm
c}v_{\rm c}l/(vl_{\rm c})$, where $v$ is the r.m.s. turbulent
velocity in the ambient medium. With $f_{\rm c}=10^{-3}$--$10^{-4}$
(Berkhuijsen 1999 and references therein), $l/l_{\rm c}\simeq10$ and
$v/v_{\rm c}\simeq2$, we obtain
$V/l^3\simeq5\times(10^{-2}$--$10^{-3})\ll1$, so the coupling between
molecular clouds and the intercloud magnetic fields is expected to be
weak. Molecular clouds move with respect to the surrounding gas, so
their connection with the external magnetic field can be lost
rapidly. Equipartition can still be maintained within the clouds but
may not occur between the bulk motion of the clouds (that provides
the average pressure support to the ISM) and the turbulent magnetic
field.

The same arguments apply to the regular magnetic field and one might
expect a relatively weak dependence of $B$ on the parameters of the
molecular gas (cf.\ Beck 1991). We show in Fig.~\ref{fig:b} the
vertical profile of the turbulent magnetic field $b$ calculated from
Eq.~(\ref{eq:energy}) for a choice of $K$ values, where the
contribution of molecular clouds to kinetic energy density has been
suppressed. With the contribution of molecular gas to $b$ suppressed
in Eq.~(\ref{eq:energy}), the turbulent magnetic field is reduced
near $z=0$ giving room for a stronger regular magnetic field. The
scale height $h_b$ is not sensitive to different values of $K$.

%-------------------------------------------------------------------
\subsection{Equilibrium near the Sun: discussion}
The comparative simplicity of the results obtained with
$B_0\simeq4\mkG$, compared to those discussed in Section~\ref{B2} for
weaker $B_0$, as well as their quantitative and qualitative
plausibility leads us to favour the arguments suggesting
$B_0\simeq4\mkG$ in the solar vicinity.

The results of our model with $B_0\simeq4\mkG$, discussed in
Section~\ref{B4}, compare reasonably well with the vertical distribution
of the magnetic field derived by Kalberla \& Kerp (1998, K\&K
hereafter) using a hydrostatic equilibrium model, shown in their
Fig.~5. In particular, K\&K predict a regular magnetic field in the
halo of $B\simeq 1\mkG$ at $z\simeq 8\kpc$ as a requirement for
hydrostatic equilibrium. However, K\&K have a weaker turbulent
magnetic field, with $b_0\simeq 3.5\mkG$, that extends to only
$z\simeq2\kpc$ compared to $z>8\kpc$ in our model. Polarized radio
emission from the haloes of the edge-on spiral galaxies NGC~891 and
NGC~4631 is significantly depolarized by turbulent magnetic fields
(Hummel et al.\ 1991) and any large scale fields present in the halo
must become tangled by turbulence and Parker and thermal
instabilities (Beck et al.\ 1996), so we expect to see a turbulent
component of the magnetic field at all $z$.

In contrast to K\&K, who have magnetic pressure in the disc of $1/3$
the total gas pressure and no pressure contribution from cosmic rays,
we find equality between the gas (turbulent plus thermal) and total
magnetic field pressures in the region $|z|<1 \kpc$. And whereas K\&K
have pressure equilibrium between gas and magnetic fields in the halo
at $z\simeq5 \kpc$ our model only requires $P_{\rm m} \simeq 0.5P_{\rm
  g}$. We have assumed equipartition between cosmic ray and magnetic
pressures at all $z$.

%--------------------------------------------------------------
\subsubsection{Implications for external galaxies}

%***FIGURE---------------------------------------------------------
\begin{figure}
\epsfxsize=8.0cm
\centerline{\epsfbox{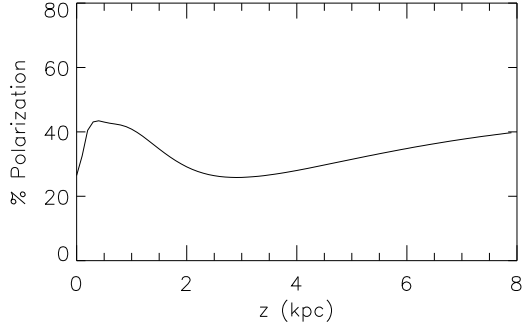}}
  \caption[]{Expected percentage polarization of non-thermal radio
emission, using the regular and turbulent magnetic field
    distributions from
    Figs~\protect\ref{fig:NewB}~and~\protect\ref{fig:Newb} (for $K=0.8$) in
    Eq.~(\ref{eq:pol}), with ${\cal P}_{0}=70\%$.}
\label{fig:Pol}
\end{figure}
%***----------------------------------------------------------------

The observed degree of linear polarization of synchrotron radio
emission is given by Burn (1966) as
\begin{equation}
  \label{eq:pol}
  {\cal P}={\cal P}_{0}\frac{B_{\perp}^2}{B_{\perp}^2+b_{\perp}^2}\;,
\end{equation}
where $B_{\perp}$ and $b_{\perp}$ are, respectively, the regular and
turbulent (isotropic) magnetic field components transverse to the line
of sight and ${\cal P}_{0}\approx70\%$ is the degree of polarization
in a purely regular field. Assuming that $b_{\perp}^2/B_{\perp}^2
\simeq b^2/B^2$ Fig.~\ref{fig:Pol} shows the expected variation in
${\cal P}$ with $z$.

An implicit assumption in Fig.~\ref{fig:Pol} is that ${\cal P}_{0}$
is constant in $z$.  However, ${\cal
P}_{0}=(\gamma+1)/(\gamma+\sfrac73)$ depends on the electron energy
spectral index $\gamma$. If $\gamma$ increases with $z$, due to
cosmic ray electron energy losses as they travel into the halo, then
${\cal P}_{0}$ and hence ${\cal P}$ will increase with $z$ making the
slope at $z\ga 3\kpc$ in Fig.~\ref{fig:Pol} even steeper, although
this effect is only weak.

Observations of polarized radio emission from edge-on galaxies, such
as NGC~891 (Hummel et al.\ 1991, Dumke et al.\ 1995) and NGC~4631
(Hummel et al.\ 1991, Golla \& Hummel 1994, Dumke et al.\ 1995) show
the degree of polarization rising with $z$ from a minimum at $z=0$.
The observed small ${\cal P}$ at the midplane, e.g.\ ${\cal
P}\vert_{z=0}\simeq1\%$ in NGC~891 (Hummel et al.\ 1991), can be
explained by strong Faraday depolarization by both regular and
turbulent fields in the disc.  Above $z\simeq2\kpc$, the steady
increase in ${\cal P}$ shown in Fig.~\ref{fig:Pol} is reproduced in
the observations, but ${\cal P}$ in Fig.~\ref{fig:Pol} is a factor of
two higher than observed in external galaxies at the wavelength
$\lambda=20\cm$. Faraday depolarization due to both regular and
turbulent magnetic fields is expected to be weak in galactic halos
where Faraday rotation measure does not exceed $10\radm$ due to low
gas density. For example, depolarization by internal Faraday
dispersion arising from a turbulent magnetic field $b\simeq2\mkG$ with
correlation length $l\simeq0.5\kpc$ (Poezd, Shukurov \& Sokoloff
1993), path length through the halo $L\simeq10\kpc$ and thermal
electron density $n_{\rm e}\simeq10^{-3}\cmcube$ can only reduce the
degree of polarization by a few percent even at $\lambda=20\cm$,
${\cal P}={\cal P}_0(1-e^{-S})/S\approx0.96{\cal P}_0$, where
$S=4\lambda^4(0.81n_{\rm e}b)^2lL$ (Burn 1966, Sokoloff et al.\
1998). 

A plausible explanation of the relatively low polarization of
synchrotron emission from galactic halos of edge-on galaxies is that
$b_\perp^2/B_\perp^2$ is larger ($\simeq6$) than predicted by our
model (2--3) for the Milky Way.

%***FIGURE---------------------------------------------------------
\begin{figure}
\epsfxsize=8.0cm
\centerline{\epsfbox{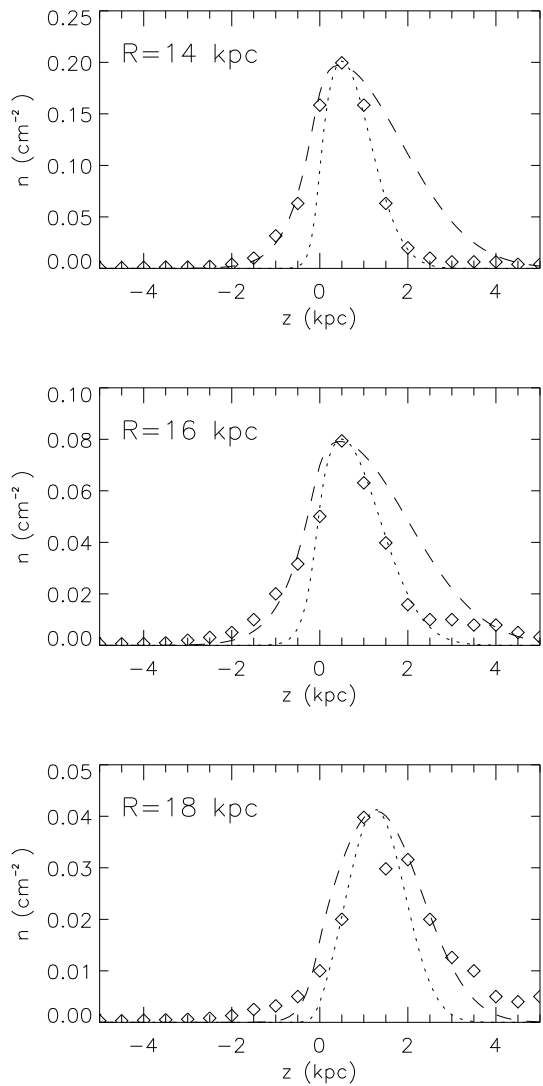}}
  \caption[]{\HI\ number
    density from Diplas \& Savage (1991) at various $R$ and
    $\Theta=90^{\circ}$ (diamonds) and fits using two different
    velocity dispersions in a warped gas disc with the asymmetric
    gravity of Fig.~\protect\ref{fig:grav}. Dashed: $u=7,$ 8 and
    $5\kms$ from top to bottom. Dotted: and $u=15,$ 15 and $8\kms$
    from top to bottom.}
\label{fig:gas}
\end{figure}
%***----------------------------------------------------------------

%****************************************************
\section{Hydrostatic equilibrium in the Outer Galaxy} \label{OG}
%****************************************************
The Galactic gas disc is warped beyond the solar orbit (Binney \&
Tremaine 1987, Binney 1992). The warp is thought to propagate through
the disc in azimuth and the speed of propagation determines whether
the ISM at large radii has enough time to reach a state of hydrostatic
equilibrium.  The period of the warping wave is about $3\times
10^8\yr$, assumed to be similar to the disc rotation period at a
radius 12\,kpc where the warp starts (Binney 1992). This is longer
than the sound crossing time over a scale height of $1\kpc$ at the
sound speed of $10\kms$.  Thus, hydrostatic equilibrium can be a valid
first approximation for the state of the gas at $R=12$--18\,kpc (see,
however, Masset \& Tagger 1996, for a discussion of possible
deviations from equilibrium).

In this section we argue that the warping can lead to a noticeable
asymmetry in the vertical profile of the gas density and confirm this
using the vertical $\HI$ distributions in the outer Galaxy obtained by
Diplas \& Savage (1991).

We describe positions in Galactocentric polar coordinates
($R,\Theta,z$), where $R$ is distance in the Galactic plane from the
Galactic centre, $\Theta$ is the azimuth in the Galactic plane, with
the Sun at $\Theta=180^\circ$ and $\Theta$ increasing anti-clockwise
viewed from the north Galactic pole; and $z$ is vertical distance from
the Galactic plane.  The galactocentric distance of the Sun is adopted
as $R_{\odot}=10\kpc$, rather than the currently used
$R_{\odot}=8.5\kpc$ to avoid the need to rescale the data of Diplas \&
Savage (1991).

%-----------------------------------------------------
\subsection{The observed vertical asymmetry in  H\,{\bf\sc i} density}
Data from the northern hemisphere \HI survey (Stark et al.\ 1991) was
used by Diplas \& Savage (1991) to study the gas morphology in the
outer Galaxy. The data show a distinct asymmetry in the vertical
distribution of \HI across a substantial region of the outer disc. The
asymmetry can also be seen, but less distinctly, in maps of the outer
Galaxy presented by Burton (1988, Fig.~7.18).

The asymmetry occurs over a sector of the outer Galactic disc in the
azimuthal range $\Theta=60^\circ$ to $\Theta=110^\circ$, visible
between $R=14$ and $18\kpc$. It is notable that the warping is maximum
in this region (Diplas \& Savage 1991). Unfortunately, the region
around $\Theta=270^\circ$ where the other crest of the warp is located
is not accessible for a northern hemisphere survey.

Diplas \& Savage (1991) do not comment directly on the asymmetry, but
the data reduction process used can lead to some mis-positioning of
gas, especially at lower densities, because of confusion with \HI
clouds moving with peculiar velocities. (B.~Savage, 1998, private
communication). Reversing the calculations of Diplas \& Savage (1991)
in the region of asymmetry shows that the velocity of the \HI, $-100 <
v < -40\kms$, is indeed similar to that of intermediate velocity
clouds (Wesselius \& Fejes 1973, Kuntz \& Danly 1996). However the
region ($35^\circ\la l \la78^\circ$ and $-8^\circ\la b \la8^\circ$,
where $l$ and $b$ are Galactic longitude and latitude) is too large to
be associated with a single cloud. We proceed with the assumption that
the asymmetry is a real feature of the outer Galaxy gas distribution.

The results of Diplas \& Savage (1991), for three positions in the
outer Galaxy, are illustrated in Fig.~\ref{fig:gas}. The maximum gas
density is displaced to positive $z$, reflecting the warp of the gas
disc. It is also clear that the gas density is asymmetric with respect
to the displaced midplane, with a density excess at positive $z$.

%***FIGURE---------------------------------------------------------
\begin{figure}
\epsfxsize=8.0cm
\centerline{\epsfbox{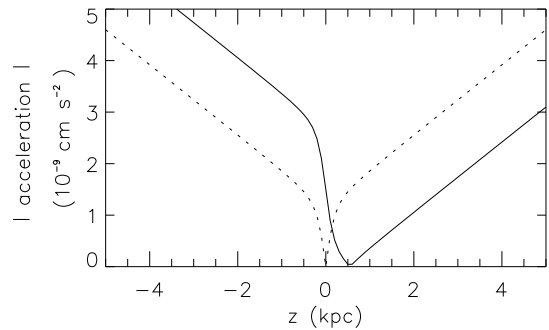}}
\caption[]{The vertical acceleration in a warped disc. Radial extra\-polation,
  to $R=16\kpc$, of the acceleration due to gravity from the solar
  position after Ferri\`ere (1998), $|g|$ from
  Eq.~(\protect\ref{outer_gravity}) (dotted) and $|g+a_0|$ with
  constant $a_0$ added to move the minimum to $z=0.5\kpc$ in a rough
  model of the force responsible for the warp (solid).}
\label{fig:grav}
\end{figure}
%***----------------------------------------------------------------

%------------------------------------------------------------------
\subsection{Asymmetry in the vertical H\,{\bf\sc i} density caused by the warp}
To model gas equilibrium in a warped disc, we describe the warp
locally as being produced by a $z$-independent acceleration $a_0$,
added across all $z$ as required to make $z=0.5\kpc$ the bottom of the
potential well. Since the stellar disc is not significantly warped, we
scaled $g$ of Eq.~(\ref{gz}) to the outer Galaxy (Ferri\`ere 1998)
(neglecting edge effects and gravity from the gas since the region of
interest is not far from the edge of the stellar disc) and then added
a constant acceleration $a_0=1.5\times10^{-9}\cms$ at $R=16\kpc$. The
modulus of the resulting vertical acceleration is shown in
Fig.~\ref{fig:grav}. Acceleration of this magnitude displaces the gas
density maximum to the observed height at this radius. Thus, the
vertical equilibrium equation (\ref{eq:equilibrium}) is replaced by
\begin{equation}
  \frac{\partial P}{\partial z}=\rho (g+a_0)\;,      \label{eq:gog}
\end{equation}
with
\begin{equation}
  -g=
  \frac{A_1z}{\sqrt{z^2+Z_1^2}}\exp\left(-\frac{R-R_{\odot}}{R_1}\right)
  +A_2\frac{z}{Z_2}\,\frac{R_{\odot}^2+R_2^2}{R^2+R_2^2}\;,
  \label{outer_gravity}
\end{equation}
where $A_1,\ A_2,\ Z_1$ and $Z_2$ are defined after Eq.~(\ref{gz})
and $R_1=4.9\kpc,\ R_2=2.2\kpc$.

If all of the pressure sources are symmetric with respect to the
shifted midplane, the asymmetric force of Eq.~(\ref{eq:gog}) should
give rise to a larger gas scale height above the shifted midplane at
$0.5 \kpc$, similar to what is observed -- see fits in
Fig.~\ref{fig:gas}.

To further, investigate the effect of the warp, we use Parker's (1966)
model of hydrostatic equilibrium
\begin{equation}
\label{eq:Parker}
  \rho(z)=\rho_0\exp\left[\frac{-\Phi(z)}{(1+\alpha+\beta)u^2}\right]\;,
\end{equation}
where $\Phi$ is the gravitational potential, obtained by integrating
Eq.~(\ref{outer_gravity}) and $\alpha$ and $\beta$ represent the
pressures due to magnetic fields and cosmic rays, respectively, as
multiples of the gas pressure. A more detailed analysis as in
Section~\ref{HE} is hardly warranted here because of our limited
knowledge of the gas parameters in the outer Galaxy. Assuming
$\alpha=\beta=1$ and using the asymmetric force of Eq.~(\ref{eq:gog}),
we fitted Eq.~(\ref{eq:Parker}) to the observed density distribution
using different values of the effective total gas velocity dispersion,
$u$.

As shown in Fig.~\ref{fig:gas}, the density distributions are
reproduced satisfactorily close to the shifted midplane with a single
(smaller) velocity dispersion. However, there is additional asymmetry
at larger heights which can be accounted for by larger values of $u$
at negative $z$, at least at $R=14$ and $16\kpc$. The same effect is
found at other positions in the sector of the Galactic disc that is
positively warped; at each position a further source of asymmetry is
required as a higher order effect at large heights.

\begin{figure}
\epsfxsize=8.0cm
\centerline{\epsfbox{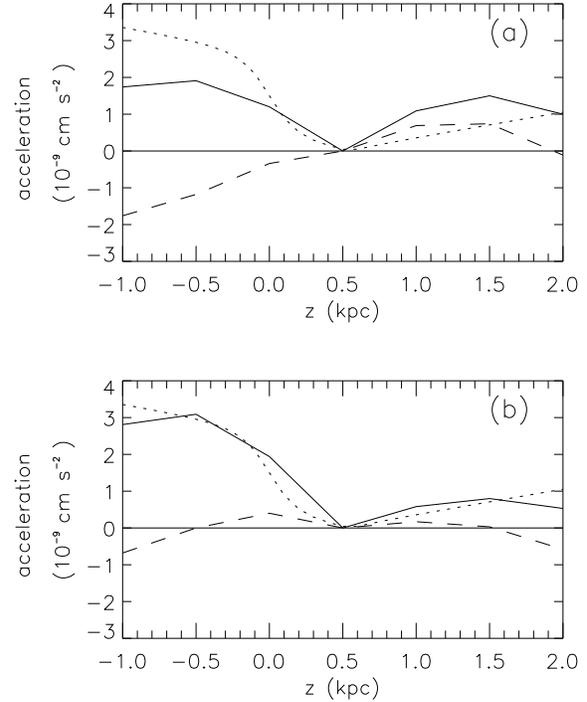}}
  \caption[]{Solid: $|\rho^{-1}\partial P/\partial z|$, the total
  vertical acceleration due to gravity and the warping force required
  to maintain the observed \HI in hydrostatic equilibrium at
  $R=16\kpc$. Dotted: the acceleration $|g+a_0|$ as in
  Fig.~\protect\ref{fig:grav}. Dashed: $a_1$, the difference between
  the above two. {\bf(a)}: the above with a single velocity
  dispersion, $u=11.5\kms$, and {\bf(b)}: a composite best fit with
  $u=8\kms$ above the shifted midplane and $u=15\kms$ below.}
  \label{fig:grav2}
\end{figure}

To quantify deviations from Eq.~(\ref{eq:gog}) with a constant $a_0$,
we show in Fig.~\ref{fig:grav2} the residual acceleration
\[
a_1=\rho^{-1}\frac{\partial P}{\partial z}-g-a_0\;,
\]
at $R=16\kpc$ both for a uniform velocity dispersion $u=11.5\kms$, a
mean between the two approximations of Fig.~\ref{fig:gas}, and for
$u=8\kms$ (shown dotted in Fig.~\ref{fig:gas}) above the shifted
midplane and $u=15\kms$ (dashed in Fig.~\ref{fig:gas}) below the
midplane. The residual, asymmetric component of the acceleration is
$|a_1|\simeq10^{-9}\cms$ or less at $\Delta z=\pm1\kpc$ from the
midplane of the warped disc, so $|a_1/a_0| < 0.7$.

%***FIGURE---------------------------------------------------------
%***----------------------------------------------------------------

The former option, illustrated in Fig.~\ref{fig:grav2}a, attributes
the whole residual asymmetry to a $z$-dependent component of the
warping force, $a_1$, which then must vary at a scale of order 1\,kpc,
with $a_1$ possibly becoming comparable to $a_0$ at large heights.

In the latter case, illustrated in Fig.~\ref{fig:grav2}b, one
supposes that a part of the residual asymmetry is due to an asymmetric
pressure in the ISM. We have quantified the residual in terms of
velocity dispersions. An obvious candidate for the residual asymmetry
can be the regular magnetic field $B(z)$. We note that some galactic
dynamo models are compatible with strongly asymmetric $B(z)$ where
$B(z)$ vanishes in a layer at a height of about $1\kpc$ on one side of
the disc. This could happen if regular magnetic fields in the disc and
halo have different parities in $z$ (Sokoloff \& Shukurov 1990,
Brandenburg et al.\ 1992). The range of $a_1$ in
Fig~\ref{fig:grav2}b, $\Delta a_1$, would correspond to an
asymmetric component of $B(z)$ of order $\sqrt{\rho\Delta a_1 \Delta
  z}\simeq1\mkG$, where $\Delta a_1\simeq a_1$, gas density
corresponds to $0.03\cmcube$, and $B$ has to be smaller above the
midplane $z>0$.

%-----------------------------------------
\section{Summary}
As discussed in Section~\ref{RMF}, parameters of a multiphase ISM in
the solar vicinity are compatible with a model of hydrostatic
equilibrium where approximate equipartition exists between the energy
density of the turbulent magnetic field and the turbulent energy
density of the gas in its diffuse phases. Models of hydrostatic equilibrium
are sensitive to the midplane strength of the regular magnetic field,
$B_0$.

If $B_0\simeq 2\mkG$, the ISM may be overpressured at $1\la
z\la4\kpc$ (so hydrostatic equilibrium is a poor approximation there),
and this can contribute to driving the galactic fountain. The
interstellar magnetic field should be coupled more weakly to the
molecular clouds than with the other, diffuse gas phases; otherwise,
the regular magnetic field strength has a deep, narrow minimum at
$z=0$.

If $B_0\simeq 4\mkG$ then our model gives two-component vertical
profiles of both regular, $B$, and turbulent, $b$, magnetic fields.
The narrower components have scale heights of $h_B\simeq1\kpc$ for the
regular magnetic field, $h_b\simeq 0.5\kpc$ for the turbulent magnetic
field. The more extended components have an approximately constant
regular magnetic field strength of $B\simeq 1\mkG$ for $2\la z \la 8
\kpc$ and $b_0/B_0=1.25$ at $z=0$. The vertical dependence of the
degree of linear polarization of synchrotron emission expected from
the derived $B(z)$ and $b(z)$, shows a steady increase above $z\simeq
2\kpc$ similar to that observed for external edge-on galaxies.

Models with $B_0\simeq4\mkG$ seem to be simpler and more attractive
than those with $B_0\simeq2\mkG$. The former are consistent with
hydrostatic equilibrium maintained at all heights, whereas the
overpressure at $1\la z\la4\kpc$ in the latter models drives
systematic vertical gas flow to the halo at a speed of about
$50\kms$. This counterintuitive conclusion results from a larger
contribution of turbulent magnetic fields and cosmic rays to the
total pressure required in the latter models for hydrostatic
equilibrium at $z\la1\kpc$. We should emphasize that our results with
$B_0\simeq2\mkG$ are only tentative at $z\ga1\kpc$ where our basic
assumption of hydrostatic equilibrium does not apply with this value
of $B_0$.

We have demonstrated in Section~\ref{OG} that the vertical force
acting on the gas in a warped disc should become asymmetric in $z$,
resulting in asymmetry in the gas distribution. By taking the simplest
form for the vertical structure of the warping force as a first
approximation and assuming hydrostatic equilibrium can be reached, we
calculated the expected vertical gas density distribution. The
observed warping is produced by an acceleration of $a_0 =
1.5\times10^{-9}\cms$ at $R=16\kpc$. The vertical profiles of neutral
hydrogen density in the outer Galaxy (Diplas \& Savage 1991) show an
asymmetry of the expected form, with the density $0.5\kpc$ above the
shifted midplane a factor of $1.2$ higher than the density $0.5\kpc$
below.

%----------------------------------------------------------------
\section*{Acknowledgements}
We are grateful to A.~Brandenburg, W.~Dobler, K.~Ferri\`ere, M.~Tagger
and E.~Zweibel for helpful discussions and to B.~Savage for useful
comments on his \HI data. The comments of the referee, R.~Beck,
greatly improved the paper. This work was supported by PPARC and the
Nuffield Foundation.

%----------------------------------------------------------------

\bsp
\label{lastpage}

\end{document}